\documentclass[%
 10pt,twoside,twocolumn,english,aps,amssymb,amsmath,amsfonts,superscriptaddress,showpacs,prb
]{revtex4-2}

\usepackage{graphicx}% Include figure files
\usepackage{dcolumn}% Align table columns on decimal point
\usepackage{bm}% bold math
\usepackage[T1]{fontenc}
\usepackage{natbib} %\bibliographystyle{prsty}
\usepackage[latin9]{inputenc}
\usepackage[active]{srcltx}
\usepackage{babel}
\usepackage{array}
\usepackage{mathrsfs,ulem,epstopdf}
\usepackage{bm}
\usepackage{wasysym}
\usepackage{color}
\usepackage[colorlinks,bookmarks=false,citecolor=blue,linkcolor=blue,urlcolor=blue]{hyperref}
\def\wn/{\,cm$^{-1}$}
\def\area/{\,cm$^{-2}$}
\def\cubic/{$_\mathrm{c}$}
\def\DM/{Dzyaloshinskii-Moriya}
\def\BTCP/{Ba(TiO)Cu$_4$(PO$_4$)$_4$}

\def\ntso{NaTiSi$_{2}$O$_{6}$}
\def\ltso{LiTiSi$_{2}$O$_{6}$}
\def\fd3m{Fd$\overline 3$m}
\def\p1bar{P$\overline 1$}
\def\c2c{C$2/c$}
\def\i41amd{I4$_{1}$/amd}
\def\t2g{$t_{2g}$}

\def\tio6{TiO$_6$}

\DeclareRobustCommand{\rchi}{{\mathpalette\irchi\relax}}
\newcommand{\irchi}[2]{\raisebox{\depth}{$#1\chi$}} % inner command, used by \rchi
\makeatletter
\@ifundefined{textcolor}{}
{%
 \definecolor{BLACK}{gray}{0}
 \definecolor{WHITE}{gray}{1}
 \definecolor{RED}{rgb}{1,0,0}
 \definecolor{GREEN}{rgb}{0,1,0}
 \definecolor{BLUE}{rgb}{0,0,1}
 \definecolor{CYAN}{cmyk}{1,0,0,0}
 \definecolor{MAGENTA}{cmyk}{0,1,0,0}
 \definecolor{YELLOW}{cmyk}{0,0,1,0}
}

\makeatother

\begin{document}

%\preprint{AIP/123-QED}

\title[Sample title]{Orbital disorder and ordering in \ntso: $^{29}$Si and $^{23}$Na  NMR Study}

\author{Ivo Heinmaa}
\email[]{ivo.heinmaa@kbfi.ee}
%\author{Enno Joon}
\author{Riho R\"{a}sta}
\affiliation{National Institute of Chemical Physics and Biophysics, Akadeemia tee 23, 12618 Tallinn, Estonia}
\author{Harlyn J. Silverstein}%\altaffiliation[Present address: ]{Currently not affiliated with any academic institution.}
\affiliation{Department of Chemistry, University of Manitoba, Winnipeg, MB, R3T 2N2, Canada}
\author{Christopher R. Wiebe}
\affiliation{Department of Chemistry, University of Manitoba, Winnipeg, MB, R3T 2N2, Canada}
\affiliation{Department of Chemistry, University of Winnipeg, Winnipeg, MB, R3B 2E9 Canada}
\author{Raivo Stern} %[0000-0002-6724-9834]
\email[]{raivo.stern@kbfi.ee}
\affiliation{National Institute of Chemical Physics and Biophysics, Akadeemia tee 23, 12618 Tallinn, Estonia}

\date{\today }% It is always \today
%\orcid{0000-0002-6724-9834}
%\date{\today}
%\received {29 January 2019; Revised 1 August 2019}
\begin{abstract}
\ntso \ is an exemplary compound, showing an orbital assisted spin-Peierls phase transition at T$_c$~=~210~K. We present the results of $^{29}$Si and $^{23}$Na NMR measurements of \ntso. The use of magic angle spinning techniques unambiguously shows that only one dynamically averaged silicon site can be seen at T~$>$~T$_c$. At cooling, the $^{29}$Si MAS NMR spectrum shows interesting changes. Immediately below T$_c$ the spectrum gets very broad. Cooling further, it shows two broad lines of unequal intensities which become narrower as the temperature decreases. Below 70~K two narrow lines have chemical shifts that are typical for diamagnetic silicates. The hyperfine field values for the two sites are \textit{H}$_{\text{hf}}^{29}$~=~7.4~kOe/$\mu_B$ and 4.9~kOe/$\mu_B$. In the paramagnetic state at high temperature, the spin-lattice relaxation of $^{29}$Si was found to be weakly temperature dependent. Below T$_c$ the Arrhenius type temperature dependence of the relaxation rate indicates an energy gap $\Delta$/k$_B$~=~1000(50)~K. In the temperature region from 120 to 300~K the relaxation rate was strongly frequency dependent. At room temperature we found a power law dependence $T_1^{-1}~\propto~\omega_L^{-0.65(5)}$.~% In the temperature region 
For 70~$<$~T~$<$~120~K the relaxation appeared to be non-exponential which we assigned to a relaxation due to fixed paramagnetic centers. Simulation of the magnetization recovery curve showed a temperature dependence of the concentration of these centers proportional to the magnetic susceptibility. The NMR spectrum of $^{23}$Na shows the line with typical shape for the central transition of a quadrupolar nucleus. A small frequency shift of $^{23}$Na resonance corresponds to a very small hyperfine field \textit{H}$_{\text{hf}}^{23}$~=~0.32~kOe/$\mu_B$~. In addition, at T~$>$~T$_c$ the $^{23}$Na spectrum shows another Lorentzian shaped resonance which we attribute to the Na sites where the quadrupolar coupling is partly averaged out by ionic motion. 	
\end{abstract}

\pacs{76.60.-k, 75.47.Lx, 75.50.Ee}

\maketitle

%=============================
\section{Introduction}
\begin{figure*}
	\begin{center}
		\includegraphics[width=0.95\textwidth]{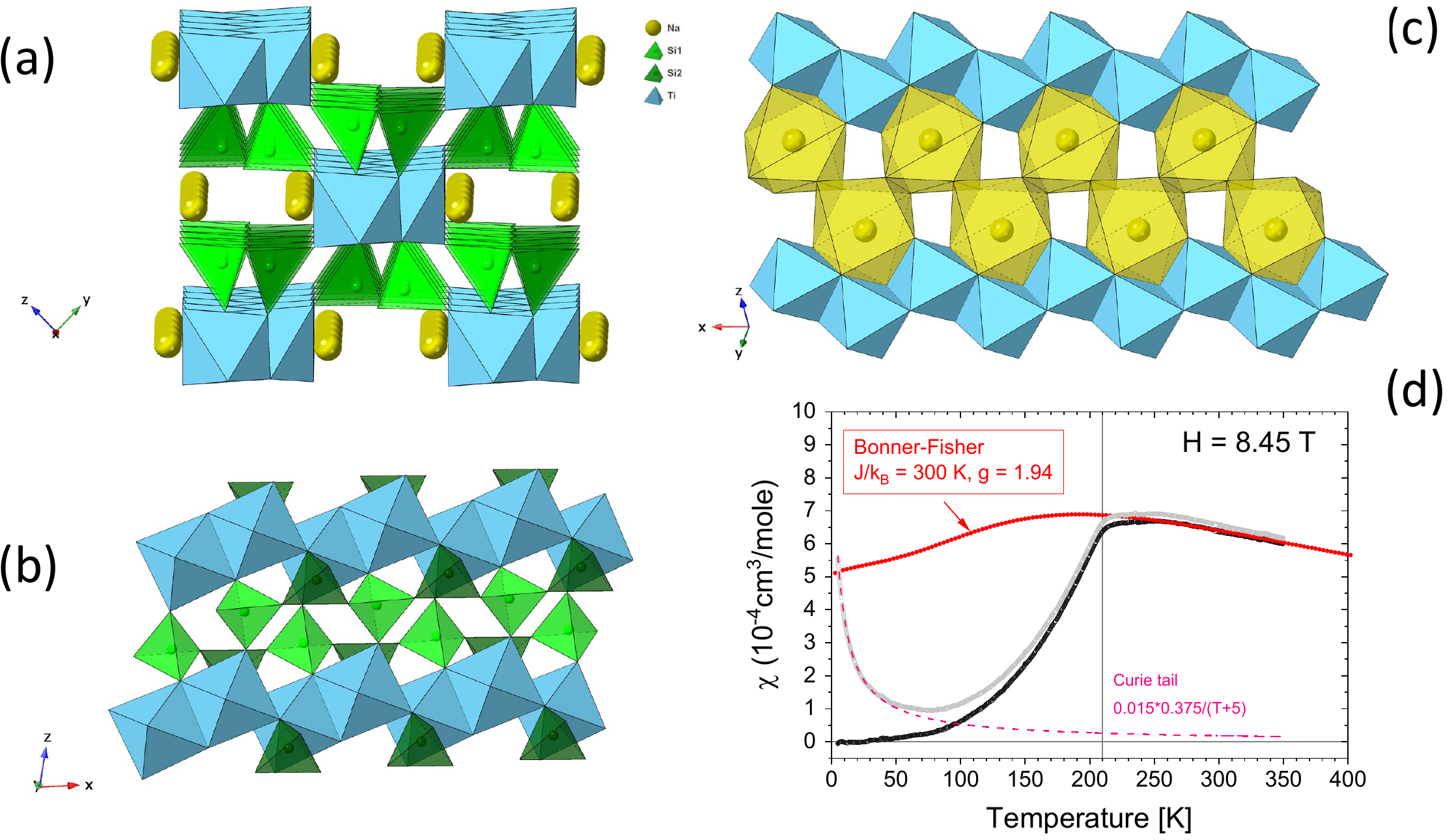} 
		\caption{\label{struc} Crystal structure of \ntso at 100~K (a, b, c)~\cite{Redhammer2003}. Edge shared TiO$_6$ octahedra (blue) form zig-zag chains which are separated by SiO$_4$ tetrahedra (light green and dark green), Na ions are forming two rows between zig-zag chains. There are two silicon sites at low temperature. In panel (b) the dimer $d_{xz}$ orbitals lay in $xz$ plane, the drawing plane. The two oxygens of Si1 tetrahedra (light green) are connected to the two neighboring Si2 tetrahedra (dark green), and the other two oxygens are connected to three Ti$^{3+}$ ions of two dimers. The Si2 tetrahedra have two nearest-neighbor Si1 tetrahedra and three nearest Ti$^{3+}$ ions, but the $d_{xz}$ orbital of the third Ti$^{3+}$ is orthogonal to the oxygen of Si2 tetrahedron, and most probably does not contribute to the spin density at Si2. Na$^+$ ions are surrounded by 8 nearest oxygen ions as shown in panel (c), four of which are connected to three nearest TiO$_6$ octahedra, four have connection to the neighboring NaO$_8$ polyhedra. The magnetic susceptibility curve of NTSO is given in panel (e). The raw data are the dark grey curve, the pink dashed line is the curve corresponding to the Curie tail ascribed to $\sim1,5\%$ localized impurities. The black line corresponds to the susceptibility after subtracting the Curie tail. The red curve corresponds to the susceptibility of homogeneous Heisenberg antiferromagnetic chain~\cite{Johnston2000} for exchange interaction between the neighbors, J/k$_B$~=~300~K, known as the Bonner-Fisher curve~\cite{BF1964}.}
		\vspace{5mm}
	\end{center}
\end{figure*}

%=============================
\setlength{\parskip}{0em}
\ntso \  (NTSO) is a Mott insulator with a clinopyroxene structure~\cite{Isobe2002,Redhammer2003} made by zig-zag chains of edge sharing \tio6 octahedra. The chains are separated by SiO$_4$ tetrahedra (FIG.~\ref{struc}). The discovery of the phase transition at T$_c$~=~210~K~\cite{Isobe2002}, where the high temperature paramagnetic spin system transforms gradually into a diamagnetic state, attracted great interest~\cite{Isobe2002,Redhammer2003,Hikihara2004,Konstantinovic2004,Baker2007,Rivas-Murias2011,Silverstein2014,Feiguin2019,Bozin2019,Streltsov2008,Streltsov2017}. For a modest spin-spin exchange coupling constant, J/k$_B$~=~300~K~\cite{Isobe2002,Silverstein2014}, T$_c$ in NTSO is very high compared to spin-Peierls transitions in other Ti$^{3+}$ spin chain compounds such as TiPO$_4$ with strong intra chain coupling J/k$_B$~=~965~K, but much lower spin-Peierls transition T$_{SP}$~=~111~K~\cite{Law2011}.% is almost 2 times lower than in \ntso. 
The powder X-ray diffraction (XRD) data showed that the magnetic phase transition in NTSO is accompanied by the structural phase transition~\cite{Isobe2002,Redhammer2003}, where the high temperature phase with equal Ti~\textendash~Ti distances transforms into the low temperature phase with alternating short/long Ti~\textendash~Ti distances. Similar phase transition at even higher temperature T$_c$~=~230~K has been detected in isostructural pyroxene compound \ltso~\cite{Isobe2002} .

The zig-zag configuration of the octahedra in NTSO chain (FIG.~\ref{struc}) puts clear constraints to the coupling between 3$d^1$ electrons of Ti$^{3+}$ ions. A single electron of the Ti$^{3+}$ shell in the crystal field of a distorted octahedron can occupy three possible \t2g orbitals: $d_{xy}$, $d_{zx}$ and $d_{yz}$. It was considered that at  high temperature, T~$>$~T$_c$, due to the equal Ti~\textendash~Ti distance, the lowest two energy levels are degenerate while the latter orbital at a bit higher energy is almost unoccupied. At low temperature the two lowest energy levels become unequal leading the system into a dimerized and diamagnetic ground state with a singlet-triplet gap $\Delta$~=~53~meV (615~K)~\cite{Silverstein2014}. 

In an early description of a 1D zig-zag chain Hikihara and Motome~\cite{Hikihara2004} pointed out two most probable configurations of Ti$^{3+}$ spin- and orbital-ordering patterns: i) the nearest neighbors have spin-ferro and orbital-antiferro configuration, and ii) the neighbors have spin-antiferro (spin-dimer) and orbital-ferro configuration. The first case favors the Hund's rule coupling and is consistent with equal Ti~\textendash~Ti distances in the chain, whereas the second case favors the super-exchange interaction, spin-singlet state of the dimers and alternating Ti~\textendash~Ti distances. Originally, the authors proposed ~\cite{Hikihara2004} that at T$_c$ the orbital order changes from the pattern i) to the pattern ii). 

{\it Ab initio} calculations of NTSO~\cite{Streltsov2006,Streltsov2008,Streltsov2017,Khomskii2021} showed that below T$_c$ the spin-dimer coupling is favored, whereas at high temperature they proposed that the $d$-electron of Ti occupies more-or-less equally all three \t2g states. Recent DFT calculations showed~\cite{Feiguin2019} that involvement of the third \t2g orbital and oxygen-atom-mediated electron hopping comparable to the direct hopping integral between neighboring Ti atoms results in much stronger quantum fluctuations. The authors calculated the energy levels and hopping integrals for several orbital configurations of zig-zag chain and found similar results: at low temperature structure the AF spin configuration is stable by about 20~meV whereas for high temperature structure the AF and FM configurations are almost degenerate. 

By their heat conductivity measurements Rivas-Murias {\it et al.}~\cite{Rivas-Murias2011} reported an interesting finding - that in the wide temperature region, 150~-~300~K, the heat conductivity  shows glass-like behavior. The behavior was attributed to the rapid orbital fluctuations above T$_c$ in an orbital-liquid state. 

Recently, the local structure of NTSO was studied by x-ray and neutron diffraction using atomic pair distribution function (PDF) method~\cite{Koch2021}. The PDF method allows to detect local interatomic distances without need to have long range order and has been used to detect local dimers in many similar dimerized compounds~\cite{Bozin2019}. Koch {\it et al.}~\cite{Koch2021} established two different Ti~\textendash~Ti distances above T$_c$ similar to the case at low T and concluded that on warming above T$_c$ the dimers evolve into a short-range orbital degeneracy lifted (ODL) state persisting up to temperature at least 490~K. Proposed ODL state involves local segments (domains) of xy-dimers and zx-dimers separated by domain walls consisting of Ti$^{3+}$ ions with antiferro-orbital couplings of occupied $yz$ orbitals or ions of occupied mixed orbitals $zx$-$xy$. 

Nuclear magnetic resonance techniques are known to provide valuable information about the local order and local dynamics of spins. In this manuscript we explore NTSO using the $^{29}$Si and $^{23}$Na NMR results. 

\begin{figure*}
	\begin{center}
		\includegraphics[width=0.95\textwidth]{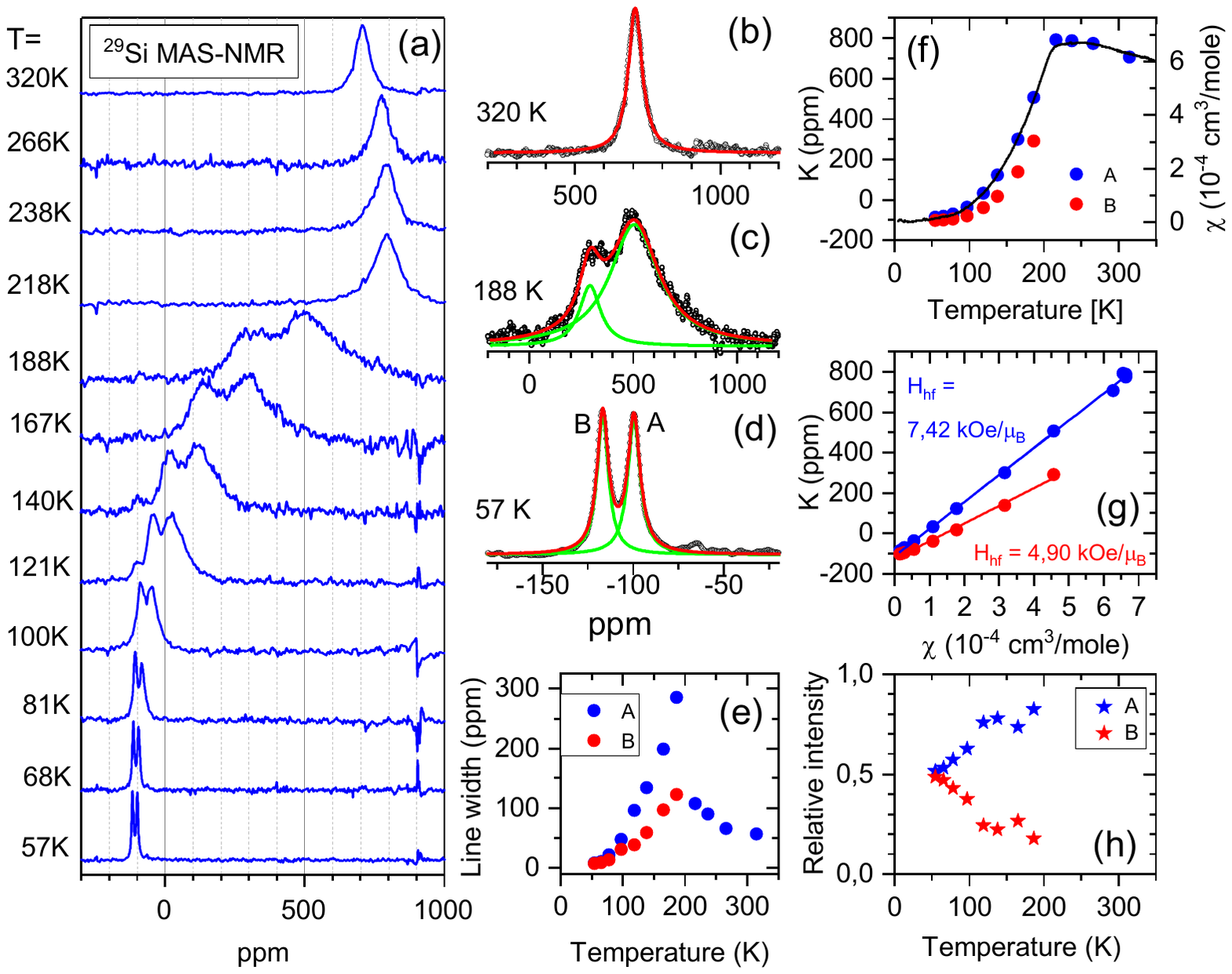} 
		\caption{\label{29Si} a) The temperature dependence of the $^{29}$Si MAS NMR spectrum of \ntso ; b) The spectrum at 320~K can be well fitted by a single Lorentzian line; c) Below the phase transition the spectrum consists of two broad Lorentzian lines with unequal intensities; d) At low temperature the spectrum contains two narrow Lorentzian lines, denoted by A and B, of equal intensities and widths; the temperature dependence of the lines width and relative intensities are given in panels (e) and (h); f) The Knight shift of both lines is proportional to the magnetic susceptibility; g) for the lines A and B two different hyperfine field values are obtained from the Clogston-Jaccarino plot~\cite{Clogston1961}.}
		\vspace{5mm}
	\end{center}
\end{figure*}

%=============================
\section{Experimental}
%=============================
A polycrystalline sample of \ntso was prepared by a solid-state reaction as described in~\cite{Silverstein2014}. The MAS NMR measurements were carried out on a Bruker AVANCE-II-360 spectrometer at 8.45~T external magnetic field (resonance frequency 71.44~MHz for $^{29}$Si and 95.119~MHz for $^{23}$Na) using home-built cryoMAS probe for 15 x 1.8 mm Si$_3$N$_4$ rotors~\cite{Topics2005,MAS-NMR_reference}. The spectra were recorded with $\pi/2-\tau-\pi-\tau-$echo pulse sequence where the delay $\tau$ was set to one rotor period. In the whole temperature region, the sample spinning frequency was adjusted to 30~kHz. The temperature of the fast-rotating sample was measured with a temperature sensor (LakeShore Cernox) at spinner assembly and corrected for given spinning frequency with earlier calibration with the temperature dependence of $^{207}$Pb chemical shift for Pb(NO$_3$)$_2$. The frequency shifts for $^{29}$Si and $^{23}$Na are given respective to TMS and NaCl resonances, respectively. The spin-lattice relaxation T$_1$  was found to be almost equal in rotating sample and static sample. Therefore, T$_1$  has been measured on static sample with saturation-recovery pulse sequence. The magnetic susceptibility was measured with PPMS (Quantum Design) VSM at 8.45 T, the same field strength as NMR spectra were recorded.

\begin{figure*}
	\begin{center}
		\includegraphics[width=0.95\textwidth]{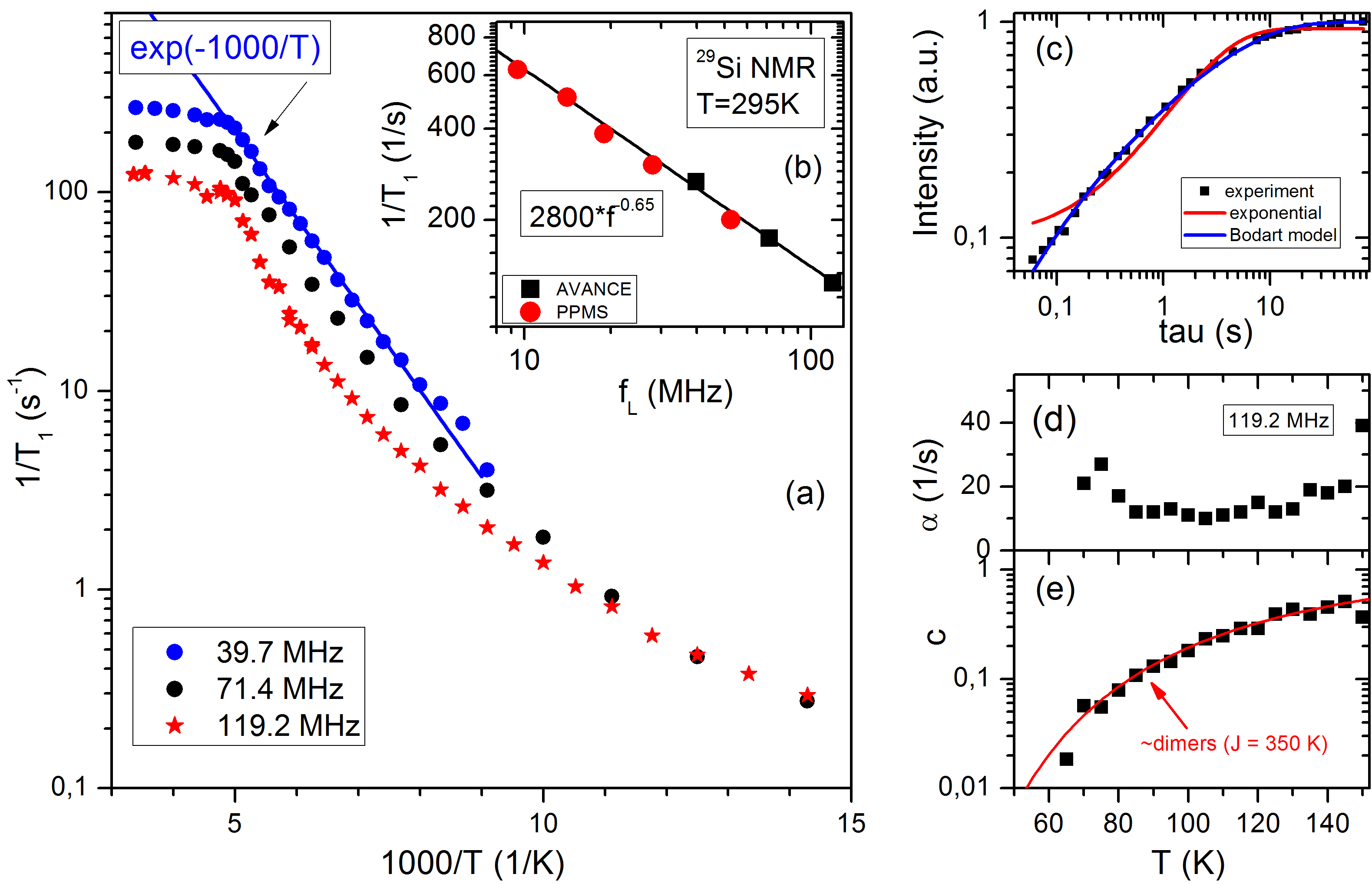} 
		\caption{\label{29SiT1} Temperature dependence of the $^{29}$Si spin-lattice relaxation rate as measured from the recovery of the intensity at maximum of the line (a). The inset (b) shows the frequency dependence of the relaxation rate at T~=~295 K. Here the data at high frequencies (black squares) are measured on AVANCE spectrometers at fixed field, the low frequency data are measured in the adjustable magnetic fields of PPMS. In panel (c) we show typical non-exponential recovery of the $^{29}$Si magnetization at temperatures below 120~K, where the red line shows exponential recovery, the blue line corresponds to the magnetization recovery due to paramagnetic centers in the model by Bodart, {\it et al.}~\cite{Bodart1996} (see text). Panel (d) shows the temperature dependence of amplitude (alpha) from the fit to the non-exponential recovery, and the panel (e) shows the temperature dependence of the concentration c from the same fit (see text). The red line in panel (e) corresponds to the temperature dependence of magnetic susceptibility of isolated dimers with exchange coupling J~=~350~K (Eqn.~4).}
		\vspace{5mm}
	\end{center}
\end{figure*}

%=============================
\section{Results}
%=============================
%=============================
\subsection{$^{29}$Si MAS NMR spectra}
%=============================
\setlength{\parskip}{0em}
Temperature dependence of the $^{29}$Si MAS NMR spectrum is given in FIG.~\ref{29Si}. At T~=~320~K the spectrum shows a single Lorentzian line at 707~ppm with a width (FWHH) of 57~ppm. With decreasing  temperature the isotropic Knight shift follows the temperature dependence of the magnetic susceptibility. At the phase transition the resonance line broadens abruptly. In the spectrum recorded at 188~K another line (line B), at smaller Knight shift is clearly seen. At lower temperatures, the spectrum shifts to the diamagnetic direction, the lines become narrower, the intensity of the line B increases and that of the line A decreases. At temperature 57~K the spectrum shows two narrow lines at -100 and -117~ppm (typical range of silicon chemical shifts of silicates~\cite{Engelhardt1987}) with the widths of 8.3 and 7.5~ppm, respectively. Following the study of the $^{29}$Si chemical shifts in titanosilicates~\cite{Balmer1997} one should expect that the  silicon with effectively two nearest neighbor Ti$^{3+}$ ions can be assigned to the line at  -117~ppm, and the line at  -110~ppm belongs to silicon with effectively three nearest Ti$^{3+}$ neighbors. The Knight shift of both lines follows linearly the magnetic susceptibility curve. 
\begin{figure*}
	\begin{center}
		\includegraphics[width=0.95\textwidth]{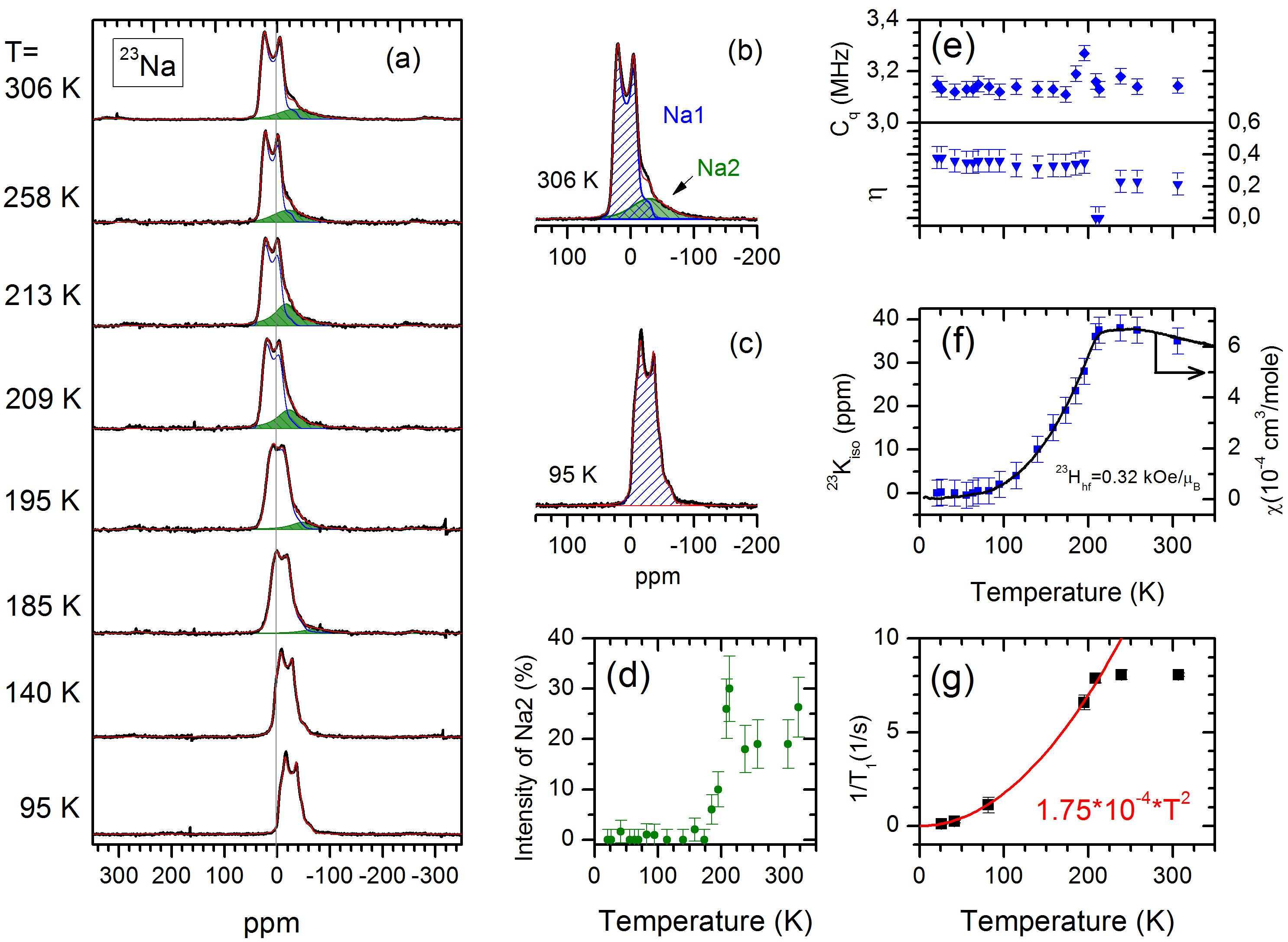} 
		\caption{\label{23Na} Temperature dependence of the $^{23}$Na MAS NMR spectrum of NTSO: (a) the black line is the recorded spectrum; the red line corresponds to the fitting. The detailed view of the spectrum at 306~K (b) shows two resonances corresponding to Na sites denoted as Na1 (blue stripes) and Na2 (green); the line corresponding to Na1 sites has typical double horn shape of central transition of quadrupolar nucleus in powder sample, whereas the resonance corresponding to the sites Na2 has featureless Gaussian/Lorentzian shape. Below the phase transition the spectrum shows only one Na1 line (c). Temperature dependence of relative intensity of the Na2 resonance (d). Panel (e) shows temperature dependence of the quadrupolar coupling constant, $C_Q$, and the asymmetry parameter, $\eta$, as fitting parameters to the Na1 line shape. Panel (f) shows that temperature dependence of the isotropic value of magnetic shift $^{23}K_{iso}$ (blue squares) scales perfectly with the temperature dependence of the magnetic susceptibility $\rchi$(full line). The hyperfine field value $^{23}H_{hf}$~=~0.32~kOe/$\mu_B$ was obtained from $^{23}K_{iso}$ vs $\rchi$ plot. Temperature dependence of sodium spin lattice relaxation rate 1/T$_1$ is given in the panel (g). }
		\vspace{5mm}
	\end{center}
\end{figure*}
\begin{equation} \label{C-J} 
K(T)=K_0+\frac{\textit{H}_{\text{hf}}}{N_A\mu_B}\rchi.
\end{equation}
\noindent
Here, $K_0$ is the temperature-independent shift, the chemical shift, and \textit{H}$_{\text{hf}}$ is the hyperfine coupling constant. 
From Clogston-Jaccarino plot~\cite{Clogston1961} (FIG.~\ref{29Si}g), K vs $\rchi$, we determined the value of hyperfine field \textit{H}$_{\text{hf}}$~=~7.42 and 4.90~kOe/$\mu_B$ for A and B resonances, respectively. 
%=============================
\subsection{$^{29}$Si Spin-Lattice relaxation} 
%=============================
$^{29}$Si spin lattice relaxation T$_1$ was measured using static sample and saturation-recovery pulse sequence. At temperatures above the phase transition, T~>~210~K, the magnetization recovery was single exponential. Due to inhomogeneous broadening at temperatures below the phase transition, the magnetization recovery at different positions of the broad line showed different T$_1$ values. The data in FIG.~\ref{29SiT1} are measured at the maximum and show almost single exponential magnetization recovery. The temperature dependence of the relaxation rate is given in the main panel of FIG.~\ref{29SiT1}. The relaxation rate at T~>~210~K is only weakly depending on temperature as expected for ordinary paramagnetic compounds. According to Moriya~\cite{Moriya}, the relaxation in paramagnetic state at high temperature can be given as:
\begin{equation}
	\frac{1}{T_1}=\frac{2\gamma ^2_N \sqrt{2\pi} S(S+1)}{3\omega_E z'} H^2_{hf},
\end{equation}
\noindent
where $\gamma_N$ is the nuclear gyromagnetic ratio, $ S $ is the nuclear spin, $H_{hf}$ is the total hyperfine field as in Eq.~\ref{C-J}, $z'=3$ is the number of the nearest paramagnetic neighbors next to the nucleus, the Heisenberg exchange frequency is given as
$\omega_E=(|J| k_B/ \hbar) \sqrt{2 z S(S+1)/3}$ (in units of rad$^{-1}$), where $z$~=~2  (in present case) is the number of nearest neighbor Ti$^{3+}$ ions, $S$ = 1/2 is the electronic spin, and $J$ is the magnitude of the exchange interaction (in Kelvins). Taking the hyperfine field value $H_{hf}$~=~7.42~kOe/$\mu_B$ from Clogston-Jaccarino plot (FIG.~\ref{29Si}g) and $|J|$~=~300~K, we obtain the high temperature relaxation rate T$_1^{-1}$~=~102~s$^{-1}$, which is a right order of magnitude as seen in FIG.~\ref{29SiT1}. An interesting finding of strong dependence of the relaxation rate on the $^{29}$Si Larmor frequency $f_L$  as T$_1^{-1}~\propto~f_L^{-2/3}$ (see FIG.~\ref{29SiT1}b) will be discussed below.

Below the phase transition the relaxation rate decreases fast, with activation type decay T$_1^{-1}$~$\propto$~$\exp(-1000/T)$. The activation type decay of the relaxation rate has been recorded in several spin-Peierls compounds~\cite{Smith1976,Itoh1995} and is caused by opening an energy gap in the spectrum of magnetic fluctuation. For the fully dimerized spin~$\frac{1}{2}$ Heisenberg AF chain the gap value $\Delta$~=~2J/k$_B$ is expected. From this we get an estimate to the exchange energy J/k$_B$~=~500~K. 

In the temperature range 70~<~T~<~140~K the nuclear magnetization recovery becomes non-exponential with typical $\propto~t^{1/2}$ time-dependence in the beginning of the recovery. Such behaviour has been assigned to the relaxation mechanism due to paramagnetic impurities~\cite{Blumberg1960}. The mechanism has been analyzed by two groups,~\cite{Bodart1996,Laboriau1996}, who provided almost identical model for the description of the process. In this model the magnetization recovery is given by two parameters  - the concentration of the relaxation centres, $c$, and the amplitude of the relaxation source $a$: 
\begin{equation}\label{param}
\frac{M_0-M_Z (t)}{M_0} =\frac{1}{1-c} [F(c^2 at)-cF(at)],
\end{equation}
where $F(x) = e^{-x} -  \sqrt{\pi x}$ erfc$(\sqrt{x})$, with the complementary error function erfc(z)~=~$\frac{2}{\sqrt{\pi}}\int_{z}^{\infty} e^{-t^2} dt$. 
It is noted that the actual meaning of the concentration $c$ and the amplitude $a$ are related to the symmetry of the structure. In addition, the formulae are valid within a certain region of concentrations - the model does not work on high concentrations, since the relaxation centers need to be diluted, and in our case, it does not work at very low concentration, since we have $\sim$~1\% of paramagnetic impurities from synthesis. We assumed the description is valid in the temperature region, where the parameter a is reasonably constant. Using this formalism, we obtained the parameters $a$ and $c$ from the fit to the non-exponential magnetization recovery curves in the temperature region from 60 to 140~K. Typical non-exponential magnetization recovery curve is given in FIG.~\ref{29SiT1}c. One can see that the model given by Eqn.~\ref{param} perfectly describes experimental curve. The temperature dependence of the parameter $c$ (proportional to the concentration of paramagnetic centers in the lattice) and the parameter $a$ are given in FIG.~\ref{29SiT1}e and \ref{29SiT1}d, respectively. FIG.~\ref{29SiT1}e shows that in noted temperature region the concentration of the paramagnetic centers decreases by more than an order of magnitude, while the parameter $a$ is almost constant in region between 60 to 140~K. Thus, we consider this temperature region valid for concentration measurements. 

It is natural to assume that the concentration of paramagnetic centers is proportional to the paramagnetic susceptibility $\rchi$. The red curve in FIG.~\ref{29SiT1}d is a curve proportional to the magnetic susceptibility curve of isolated dimers (see below, Eqn.~\ref{johnston}). The comparison provides an independent estimate to the value of exchange interaction energy J/k$_B$ ~=~350~`K. 

%=============================
\subsection{$^{23}$Na MAS NMR} %in BaTCPO}
%=============================
In the structure of NTSO the sodium ions have 8 oxygen nearest neighbors. Since the local symmetry is not cubic, the electric field gradient (efg) at $^{23}$Na nuclei (spin I=3/2) splits the spectrum into 3 components corresponding to the transitions between the eigenstates called satellite transitions (m~=~$\pm\frac{3}{2} \longleftrightarrow \pm\frac{1}{2}$), and central transition (m~=~$\pm\frac{1}{2} \longleftrightarrow \mp\frac{1}{2}$). In the present case the resonance of satellite lines in powder sample are too broad ($\pm~C_q$, where $C_q$ is the quadrupolar coupling constant) to be observed, but the central transition can be easily measured by MAS NMR, although the resonance line has peculiar double horn shape depending on the quadrupolar coupling constant and the asymmetry of the efg tensor. The temperature dependence of the $^{23}$Na MAS NMR spectrum is given in FIG.~\ref{23Na}a. 

Simulation of the line shape gives the values of the Knight shift, the quadrupolar coupling constant $C_q$ and the asymmetry $\eta$ of the efg tensor. The temperature dependence of the Knight shift and the quadrupolar coupling parameters are given in FIG.~\ref{23Na}e and \ref{23Na}f. At high temperature we detected another, featureless resonance line with considerable relative intensity of $\sim~25\%$ (FIG.~\ref{23Na}b). This line we assigned to a sodium ion where the quadrupolar coupling is partially averaged due to fluctuation of the efg. Below T$_c$ this line disappears (FIG.~\ref{23Na}c and \ref{23Na}d). Spin lattice relaxation of $^{23}$Na (FIG.~\ref{23Na}g) shows weak temperature dependence at T~>~T$_c$ due to fluctuations of paramagnetic spins, like the relaxation of $^{29}$Si. At lower temperatures the relaxation follows 1/T$_1~\propto$~T$^2$ dependence as typical for quadrupolar nuclei relaxing due to Raman mechanism of lattice vibrations. 

%=============================
\section{Discussion}
%=============================
%=============================
\subsection{Orbital disorder below T$_c$}
%=============================
\begin{figure}[!]
	\begin{center}
		\includegraphics[width=0.45\textwidth]{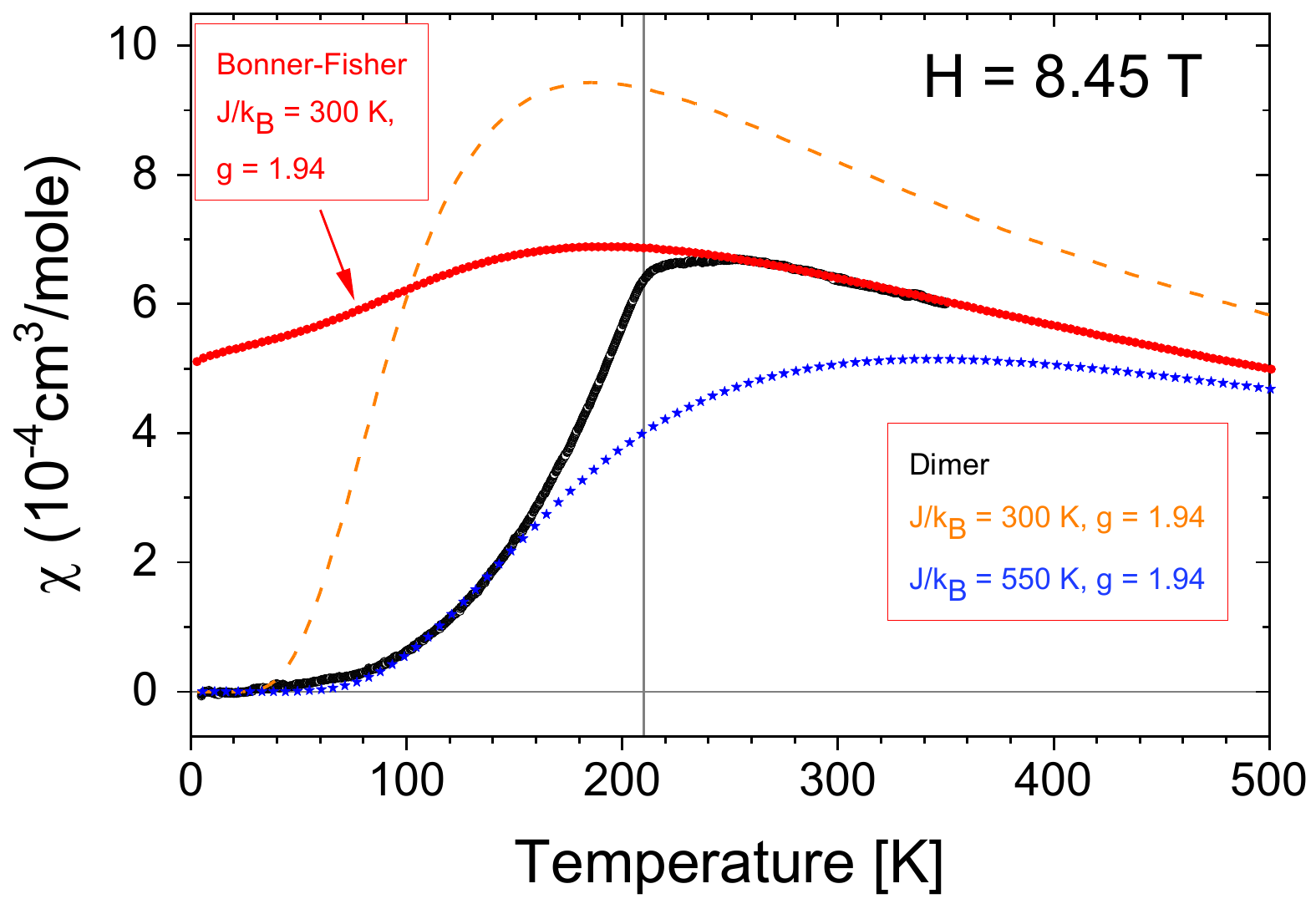} 
		\caption{\label{susc} Temperature dependence of the magnetic susceptibility of NTSO. The experimental curve with subtracted curie tail (black circles) is compared to the 							susceptibility calculated in the Bonner-Fischer model for uniform Heisenberg antiferromagnetic chain (red circles) and the susceptibility curve for isolated 							dimers with coupling constant J/k$_B$~=~300~K (dashed orange curve) and with J/k$_B$~=~550~K (blue stars). }
		\vspace{5mm}
	\end{center}
\end{figure}

The single resonance line in $^{29}$Si MAS NMR (FIG.~\ref{29Si}) of NTSO indicates that i) there is only one silicon site in the lattice, or ii) the different surroundings of silicon are averaged by fast orbital fluctuations. The case i) is clearly in contradiction with recent observations by x-ray PDF experiments~\cite{Koch2021} which clearly established different Ti~\textendash~Ti distances at temperature far above T$_c$. Therefore, the single resonance of $^{29}$Si must be a result of motional narrowing~\cite{GutoPake1950}. Taking the full linewidth in rigid lattice as 300~ppm (=~21.3~kHz) equal to the value immediately below T$_c$ (FIG.~\ref{29Si}e), we get an estimate to the correlation time of orbital fluctuations in high temperature structure as $\tau_c$~ <~10$^{-4}$~s$^{-1}$. 

According to the temperature development of the NMR spectra, below T$_c$ this fluctuation freezes, and the spectrum shows static disorder, a distribution of magnetic shifts in rigid lattice. The ordering gradually develops with decreasing temperature. At around T~$\sim$~130~K the two different silicon sites are well seen, thus we can consider this temperature as the temperature where the low temperature structure is established. At cooling further, the NMR lines become narrower as the number of paramagnetic Ti$^{3+}$ ions decreases. Dynamic/static orbital disorder in the temperature region 130~K~<~T ~ <~300~K is in good agreement with glass-like behavior of heat conductivity in this temperature region~\cite{Rivas-Murias2011}. 

The assignment of the lines A and B in the $^{29}$Si NMR low temperature spectrum to the two silicon sites in the low temperature structure of NTSO  (FIG.~\ref{struc}) can be done as given above:  the line A belongs to the site Si1, where the silicon is effectively connected to  three Ti$^{3+}$ ions; and the line B to the site Si2, connected effectively to two nearest neighbor Ti$^{3+}$ ions. This assignment is validated by the 3:2 ratio of the hyperfine fields: $^{29}H_{hf}$(Si1)~=~7.42~kOe/$\mu_B$ $vs$ $^{29}H_{hf}$(Si2)~=~4.90~kOe/$\mu_B$.

A question remains - why is the intensity of the B line at high temperatures much smaller compared to the A line? One possible way to explain this is to assume occupation of the third $d$~-~orbital, $e.~g.$ the $d_{yz}$ orbital involved in the configuration of the ODL state~\cite{Koch2021} .

%=============================
\subsection{Magnetic susceptibility}
%=============================
In FIG.~\ref{susc} we show the susceptibility curve in comparison with calculated susceptibilities. As noted above, the high temperature susceptibility is very well described by the calculation within the Bonner-Fisher model~\cite{BF1964,Johnston2000} in good agreement with the previous measurement~\cite{ Isobe2002}. Here we used the coupling parameter J/k$_B$~=~300~K and the g~=~1.94, a little lower than g = 2 (typical for $d^1$ electrons of Ti$^{3+}$ ions in octahedral field~\cite{Gerritsen1963}). The susceptibility for isolated dimers in Heisenberg AF chains has well-known analytical temperature curve for dimers~\cite{Johnston2000}: 
\begin{equation}\label{johnston}
\rchi~(T)~=~\frac{N (\mu_Bg)^2}{k_B T(3+\exp(\frac{J}{k_BT}))} %=\frac{1}{1-c} [F(c^2 at)-cF(at)],
\end{equation}
where $N$ is the Avogadro's number, $\mu_B$ is the Bohr magneton, $g$ is the electron $g$-factor and $J$ is the intradimer exchange coupling constant. If we use the same exchange coupling constant J/k$_B$~=~300~K in a dimer model, we reach to enhanced susceptibility as given by the dashed orange line in FIG.~\ref{susc}. In other words, at T$_c$ one expects the magnetization jump up (!). According to the NMR spectra, the low temperature dimerized structure is reached below T~<~130~K, where one can expect the susceptibility curve for isolated dimers is justified. Blue curve in FIG.~\ref{susc} shows good fit to the experimental data using considerably higher J/k$_B$ value of 550~K. Rather strong intra-dimer exchange couplings J/k$_B$ between 396~K to 626~K have been reported by first principles calculations~\cite{Streltsov2006,Streltsov2008} and are related to short intradimer distance d$_{Ti-Ti}$~=~3.08~\AA \ in NTSO~\cite{Redhammer2003}, much shorter than that in dimerized TiPO$_4$, with d$_{Ti-Ti}$~=~3.13~\AA \  ~\cite{Bykov2013}, where the exchange coupling, obtained from high temperature susceptibility curve, is even larger J/k$_B$~=~965~K~\cite{Law2011}. 

%=============================
\subsection{Spin-Lattice relaxation}
%=============================
An interesting finding from  $^{29}$Si spin-lattice relaxation measurements is the frequency dependence of the relaxation rate presented in~FIG.~\ref{29SiT1}b. The room temperature data show  T$_1^{-1}~\propto~f_L^{-0.65}$, where $f_L$ is the Larmor frequency. This trend was observed in the broad Larmor frequency region from 9 to 119~MHz. The temperature curves in~FIG.~\ref{29SiT1}a show that similar trend of the frequency dependence of T$_1$ is present at temperatures down to~$\sim$~130~K. Recall, in ordinary 3D paramagnetic compounds the nuclear relaxation is independent of the Larmor frequency~\cite{Moriya} depending only on the exchange frequency of electron spins as given above. 

In 1D AF Heisenberg chains the relaxation at low frequencies is enhanced due to 1D spin diffusion. The spin diffusion causes a low-frequency enhancement of the spectral density of spin fluctuations and is observed in the NMR relaxation as 1/T$_1$~=~A~+~B~$\omega_L^\frac{1}{2}$, where A is the relaxation due to the spin exchange as given above, and B is a constant (see e. g. Refs~\cite{Borsa1974,Boucher1976,Takigawa1996,Adelnia2015}). In some cases the dependence is weaker. For example, careful muon spin relaxation $\lambda$ measurements as a function of magnetic field $B$ on an organic radical-ion salt with ideal S = 1/2 Heisenberg AF chain showed field (frequency) dependence as~$\lambda~\propto~B^{-n}$ with $n$~=~0.350(7)~\cite{Pratt2006}, in reasonable agreement with theory~\cite{Muller1988}. 

We are not aware of nuclear relaxation measurements on AF chains with orbital disorder, but it is known that any kind of disorder causes enhancement of the low frequency spectral density as proved by proton relaxation study of spin dynamics in antiferromagnetic heterometallic molecular rings~\cite{Amiri2010}, where the authors demonstrate that in the homo-metallic ring Cr$_8$, the relaxation rate at high temperature was frequency independent, whereas in the substituted rings Cr$_7$Cd, Cr$_7$Ni and Cr$_7$Fe the high temperature relaxation became frequency dependent. It is interesting to note that "disorder" in series of doped paratacamites ("Herbertsmithites") MCu$_3$(OH)$_6$Cl$_2$ (M = Zn, Mg), quantum spin liquid candidates with frustrated kagome lattice, causes frequency dependence of muon spin relaxation rate 1/T$_1~\propto~\omega_L^{-0.66}$~\cite{Helton2010,Kermarrec2011}, similar to the present case.

%=============================
\section{Conclusion}
%=============================
Using $^{29}$Si and $^{23}$Na MAS NMR spectra we have studied the orbital ordering in NTSO. Temperature dependence of the spectra show that dynamic order at high temperature freezes at T$_c$ into a disordered state with different local configurations of orbitals. With cooling further, the orbitals order gradually into well defined dimers with singlet ground state. The magnetic susceptibility and spin lattice relaxation data are in accordance with relatively strong intra-dimer exchange coupling at low temperatures J/k$_B$~=~500 (50)~K. The spin-lattice relaxation rate 1/T$_1$ was found to be strongly dependent on resonance frequency. At room temperature we obtained T$_1^{-1}~\propto~f_L^{-2/3}$ in a broad frequency region.

%=============================
\section*{Acknowledgemnts}
%=============================
This research was supported by the Estonian Science Council Grants IUT23-7 and PRG4, the European Regional Development Fund Grants TK133 and TK134. 
CRW would like to acknowledge support through NSERC (CRC and DG programs. HJS acknowledges support through NSERC.

\end{document}